\documentclass[aps,prd,twocolumn,epsf,superscriptaddress]{revtex4}

\usepackage{amsfonts}
\usepackage{amsmath}
\usepackage{amssymb,amsbsy}
\usepackage{bm}
\usepackage{makeidx}
\usepackage{latexsym}
\usepackage{graphicx}
\usepackage{epsfig}
\usepackage{subfigure}
\usepackage{dsfont}
\usepackage{mathcomp}
\usepackage{color}
\usepackage{mathrsfs}
\usepackage{epsf}
\usepackage{pstricks}
\newcommand{\beq}{\begin{eqnarray}}
\newcommand{\eeq}{\end{eqnarray}}
\newcommand{\be}{\begin{equation}}
\newcommand{\ee}{\end{equation}}

\newcommand{\gapp}{\mathrel{\raise.3ex\hbox{$>$}\mkern-14mu
              \lower0.6ex\hbox{$\sim$}}}
\newcommand{\lapp}{\mathrel{\raise.3ex\hbox{$<$}\mkern-14mu
              \lower0.6ex\hbox{$\sim$}}}



\begin{document}
\title{Constraining the Equation of State of Dark Energy with Gamma Rays}
\author{Malcolm Fairbairn}

\affiliation{Physics, Kings College London, Strand, London WC2R 2LS, UK}

\begin{abstract}
\noindent
Starlight in the Universe impedes the passage of high energy (e.g. TeV) gamma rays due to positron-electron pair production.  The history of this stellar radiation field depends upon observations of star formation rate which themselves can only be interpreted in the context of a particular cosmology.  For different equations of state of dark energy, the star formation rate data suggests a different density of stellar photons at a particular redshift and a different probability of arrival of gamma rays from distant sources.  In this work we aim to show that this effect can be used to constrain the equation of state of dark energy.  The current work is a proof of concept and we outline the steps that would have to be taken to place the method in a rigorous statistical framework which could then be combined with other more mature methods such as fitting supernova luminosity distances.
\end{abstract}

\maketitle

\section{Introduction}

Observations of type 1a supernovae \cite{snorig,union2}, galaxy surveys \cite{sdss} and the CMB \cite{wmap} all seem to suggest that the expansion of the Universe is accelerating.  This fact is usually explained by the presence of some component of energy density called dark energy which has an equation of state $w\le -1/3$.  The precise equation of state of this energy component is unknown and cannot be completely determined by the data (see e.g. \cite{fairbairngoobar}).  The density of the dark energy component may be a constant, it may be growing or it may be shrinking.  Addition of the information on angular distances provided by the baryon acoustic peak which can be seen in both the two point function of luminous red galaxies at $z\sim 0.3$ and in the two point function of CMB temperature observed at redshift $z\sim 1100$ constrains the nature of dark energy quite well if the equation of state is a constant \cite{voidpaper1}, but if the equation of state $w$ is varying with cosmological epoch, the constraints on the nature of $w(z)$ are less certain and exotic equations of states where the energy density is increasing are still compatible with the data (although difficult to motivate from theories of particle physics).  

In this work, we intend to constrain the equation of state of dark energy by calculating the effect of the different corresponding expansion histories upon the opacity of the Universe to high energy gamma rays.

Stars are formed continuously throughout the history of the Universe.  Some of the first stars (the ones with low masses) are still burning after order $10^{10}$ years, whereas stars with masses larger than 10$M_\odot$ last less than a few million years on the main sequence.  The larger stars are more short lived, but much more luminous and also bluer.  The continual life cycle of stars throughout the history of the Universe creates a background radiation in addition to the CMB.  There are far fewer stellar photons than CMB photons but they have a higher energy:  The initial photons from stars range from the near infrared to the ultra violet while a lot of the starlight fails to exit the dust in galaxies and is re-emitted in the infrared.   This Extragalactic Background (star)Light or EBL therefore creates a spectrum of diffuse radiation ranging in frequency from the ultra-violet down to the border of the infrared and microwave bands at which point the CMB spectrum starts to dominate.

CMB photons have energies of about $10^{-4}$ eV so photons with energies around $10^{16}$ eV (i.e. $10^4$ TeV) will be able to pair produce electron-positron pairs upon scattering with the CMB \cite{gould}.  This sets the mean free path for such highly energetic photons to be very small in cosmological terms, only around 10 kpc.  We certainly therefore would never expect to see any such photons arriving from extragalactic sources and indeed photons with such high energies have never been positively identified (they could in principle form some fraction of ultra high energy cosmic rays but the constraints on the photon fraction is rather tightly constrained \cite{photonfraction}.  The situation could be more interesting if axion like particles exist \cite{troitsky})

Conversely, TeV energy gamma rays photons {\it have} been observed, and their arrival direction has been associated with various extragalactic sources, including some at truly cosmological distances, (e.g. 3C 279 at a redhsift of $z=0.536$) \cite{hess,magic}.  This is not completely surprising since although TeV photons can instigate electron-positron pair production upon interacting with more energetic starlight photons in the EBL, the number density of photons from stars is much lower than those in the CMB \cite{gould}.  By observing distortions in power law spectra from Blazars, several Cerenkov Telescopes have been able to place constraints upon the opacity of the Universe \cite{hess,magic}.  This tallies well with observed star formation rates and the standard cosmological model $\Lambda$CDM.  However, if one assumes that the expansion history of the Universe is that corresponding to a varying equation of state for the dark energy component, the relationship between the observations of star formation made at different redshifts and time is altered in such a way that the density of photons per square centimetre along the path of an incoming TeV gamma ray changes.  Some equations of state make the Universe less opaque, some more.  We will use the observed opacity of the Universe to place constraints on expansion histories and show that this study can rule out some cosmologies which are otherwise consistent with the data.

This extends our previous work looking at constraints upon Void universe cosmologies, cosmologies where the apparent acceleration of the Universe is due to our being located in a large underdensity \cite{voidgamma}.  Such models are already very tightly constrained by observations, so our finding a new way of putting them under pressure using gamma rays is perhaps not of interest to everyone.  Conversely, dark energy with a varying equation of state is completely consistent with all available data sets, so any new constraints will be of interest and we hope to present one in this paper.

First we will explain how we calculate the extra galactic background light from a given star formation rate, then we will describe how to calculate the opacity that light creates for high energy gamma rays.  Then we will explain how we fit the star formation rate for different cosmologies before describing the procedure we use to obtain our results.  After having presented our results we will become self critical and point out the various improvements that need to be made to the different parts of the anaylsis in order for the method to become more rigorous and useful before concluding.  The current work is therefore meant to be a proof of concept rather than providing a constraint which can be trusted at this stage as much as the standard fits.  However we hope to convince the reader that the method outlined in this work could without too much work become a new tool for understanding the nature of dark energy.

\section{The Extragalactic Background Light\label{ebl}}

\begin{figure}[t]
\includegraphics[scale=0.35]{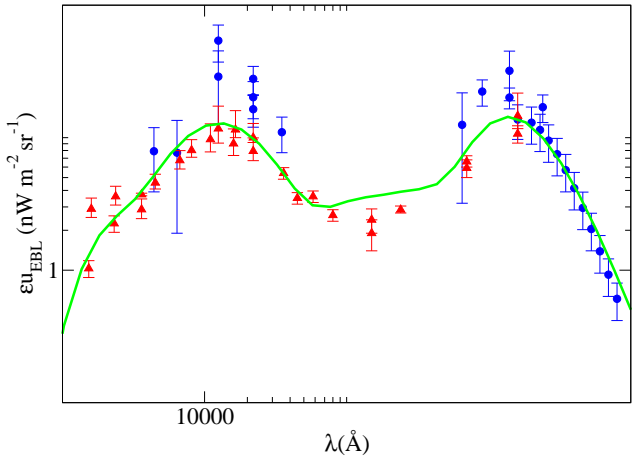}
\caption{\it The EBL spectrum produced by our code vs. the data for the $\Lambda$CDM case (solid green curve).  This is plotted against actual data where the EBL light has been detected (blue circles) and other constraints which only place lower limits (green triangles).  References for the data points are set out in the caption of figure 4 of \cite{primack}.\label{spectrum}}
\end{figure}

Because the extra-galactic background light is made of starlight, and we are surrounded by stars, it is quite difficult to see.  Nevertheless, there are lower limits on its magnitude across the whole range of frequencies due to galaxy counts and positive detections from FIRAS in the infrared (for references, see the caption of figure \ref{spectrum}).  It is possible to reconstruct the evolution and history of this radiation field using the star formation history of the Universe, which has been observed at various wavelengths \cite{sfr}.  

We do this following closely the simplified approach of Finke et al. \cite{finke} with a particular numerical implementation which lends itself to different expansion histories.  Let us outline our way of implementing those equations. The photon density of a blackbody is
\be
n_*(\epsilon,m,t_*)=\frac{dN}{d\epsilon dV}=\frac{8\pi}{\lambda_C^3}\frac{\epsilon^2}{\exp\left[\epsilon/\Theta(m,t_*)\right]-1}
\ee
where $\epsilon=h\nu/m_ec^2$ is the dimensionless photon energy in units of electron mass.  The dimensionless temperature is $\Theta(m,t_*)=k_BT(m,t_*)/m_ec^2$ where $T(m,t_*)$ is the temperature of a star of mass $m$ and age $t_*$.  We are going to assume that this is single valued, i.e. we are going to neglect the effect of metallicity variation over time.   We want to know how many photons of energy $\epsilon$ are emitted per unit time from a star of mass $m$ and age $t_*$
\be
\dot{N}(\epsilon,m,t_*)=\frac{dN}{d\epsilon dt}=4\pi R(m,t_*)^2cn_*(\epsilon,m,t_*)
\ee
where $R(m,t_*)$ is the radius of a star of mass $m$ and age $t_*$ and $c$ is the speed of light.  We obtain the temperature $\Theta$ and radius $R$ of a star of mass $m$ and age $t_*$ we follow Finke et al. in using the results from the paper by Eggleton, Fitchett and Tout \cite{eggleton} as well as the corrections outlined in the Finke work \cite{finke}.  For a population of stars of age $t_*$ but with different masses, we need a weighted integral over the range of masses to include the effects of all the different stars.  We therefore define the specific photon generation rate (per unit solar mass)
\be
\tilde{\dot{N}}(\epsilon,t_*)=\frac{dN}{d\epsilon dt}=f(\epsilon)\int_{m_{min}}^{m_{max}}\xi(m)\dot{N}(\epsilon,m,t_*)dm
\ee
where $\xi(m)$ is the initial mass function (IMF) of the stellar population.  We will assume that this also doesn't vary over time.  We assume a Salpeter initial mass function for the stars that are produced, namely $\xi(m)\propto m^{-2.35}$ with stars being produced across a range of $m_{min}=0.1 M_\odot\le m\le m_{max}=100 M_\odot$.  Here we have also introduced the effects of dimming, the parameter $f(\epsilon)$ is the fraction of photons which do not make it out of galaxies from the blue end of the spectrum due to dust.  The function $f(\epsilon)$ that we have adopted is the one set out in \cite{finke} although we have tried other dimming choices without changing the conclusions too much.

The absorbed radiation is re-emitted in the infra red and the total integrated luminosity of this dust per solar mass of a stellar population of age $t_*$ is
\be
\tilde{L}_{d}(t_*)=m_e c^2\int_0^{\infty}\left(\frac{1}{f(\epsilon)}-1\right)\epsilon \dot{N}(\epsilon,t_*)d\epsilon
\ee
The energy emitted per solar mass of a particular stellar population created at time $t_{form}$ over its lifetime up to the time of the observer (e.g. today) $t_0$ is
\begin{eqnarray}
&&\tilde{j}_{*}(\epsilon,t_0,t_{form})=\nonumber\\
&&m_e c^2\epsilon\int_{t_{form}}^{t_0}\frac{\tilde{\dot{N}}[(1+z(t))\epsilon,t-t_{form}]}{(1+z(t))}dt
\end{eqnarray}
The dust re-radiates the absorbed energy thermally.  There are at least three populations of different kinds of dust re-emitting the absorbed radiation at different temperatures so that we have
\begin{eqnarray}
&&\tilde{j}_{d}(\epsilon,t_0,t_{form})=\int_{t_{form}}^{t_0}\sum_{n=1}^3\frac{15 f_n \tilde{L}_{d}(t-t_{form})}{\pi^4\Theta_{d(n)}^4(1+z(t))}\nonumber\\
&&\times \frac{[(1+z(t))\epsilon]^3}{\exp[(1+z(t))\epsilon/\Theta_{d(n)}]-1}dt
\end{eqnarray}

The three fractions and temperatures corresponding to the three different populations of dust are fitted by hand and the good fit values we obtain are listed in table \ref{temps}.  The precise values of these numbers is not critical as it is in general the blue part of the spectrum which is responsible for the dimming of the gamma ray photons, in other words as far as opcity is concerned, the magnitude of the dimming of the bluest photons is more critical than the frequency at which they are re-emitted in the infrared.
\begin{table}
\begin{center}
\begin{tabular}{|c|c|c|}
\hline
 n & $\Theta_{d(n)}$ & $f_n$\\
\hline
1 & $7\times 10^{-9}$ & 0.7 \\
2& $25\times 10^{-9}$ & $0.15$ \\
3 & $76 \times 10^{-9}$ & $0.15$\\
 \hline
\end{tabular} 
\caption{Values for the fraction and temperature of the three populations of dust responsible for re-emitting the absorbed radiation which does not escape the galaxies.\label{temps}}
\end{center}
\end{table} 
The overall spectrum of energy density at $t_0$ is finally given by
\be
\epsilon u_{EBL}(\epsilon,t_0)=\epsilon \int_{t_{start}}^{t_0}\dot{\rho}[z(t)]\left\{\tilde{j}_{*}(\epsilon,t_0,t)+\tilde{j}_{d}(\epsilon,t_0,t)\right\}dt
\ee
where $\dot{\rho}(z)$ is the star formation rate in units of solar masses per unit time per unit volume.  To compare with observations, the energy density can be converted into intensity in units of nW m$^{-2}$sr$^{-1}$
\be
\epsilon I(\epsilon,t_0)=\frac{c}{4\pi}\epsilon u_{EBL}(\epsilon,t_0).
\ee
and the results of this spectrum can be seen for typical parameters in figure \ref{spectrum}.  
This therefore is the method by which the extragalactic background radiation is calaulated as a function of redshift.  Next we need to see how this radiation field creates a finite mean free path for high energy gamma rays.

\section{Scattering of High Energy Gamma Rays with Starlight\label{opac}}

To calculate the opacity of the Universe to photons we use the expression for the inverse mean free path (see e.g. \cite{protheroe})
\be
\lambda_{\gamma\gamma}(E_\gamma)^{-1}=\frac{1}{8E_{gamma}^2}\int_{\epsilon_{min}}^\infty \frac{n(\epsilon)}{\epsilon^2}\int_{s_{min}}^{s_{max}}s\sigma(s)ds d\epsilon
\ee
where $s_{min}=(2m_e c^2)^2$, $\epsilon_{min}=s_{min}/(4E_\gamma)$ and $\sigma(s)$ is the cross section for electron-positron pair production
\be
\sigma(s)=\pi r_e^2 \left\{(1-\beta^2)(3-4\beta^4)\ln\left[\frac{1+\beta}{1-\beta}\right]-2\beta(2-\beta^2)\right\}
\ee
where $y=s/m_e^2c^4$ is the dimensionless centre of mass energy, $\beta=\sqrt{1-2/y}$ and $\sigma_T$ is the Thomson cross section.

We then calculate the opacity as a function of redshift by integrating
\be
\frac{d\tau}{dz}=\frac{-c}{\lambda_{\gamma\gamma}(1+z)H(z)}
\ee
then the attenuation in photons (i.e. the probability of them getting through from a given time $t$) is then given by $P=\exp(-\tau(t))$.

\section{Fitting the Star Formation Rate}

The star formation rate has been observed as a function of redshift in the form of observations of galaxies at various redshifts \cite{sfr}.  In order to interpret these results in terms of a particular cosmology, one has to make a rescaling which reflects the different luminosity distances between the new cosmology and the reference cosmology in which the results are reported \cite{sfr}
\be
\dot{\rho}_*\propto\frac{L(z)}{V_c(z,\Delta z)}\propto\frac{d_C^2(z)}{d_C^3(z+\Delta z)-d_C^3(z-\Delta z)}.
\label{rescale}
\ee
In this equation, the second term reflects the fact that the star formation rate is proprotional to the inverse of the comoving volume $V_c(z)$ corresponsding to the redshift bin between $z+\Delta z$ and $z-\Delta z$ while the second term is obtained taking into account the proportionality between Luminosity and comoving distance squared $L\propto d_C^2$.

Having re-scaled the star formation rate data and its errors bars we then fit it with the ansatz
\begin{equation}
\dot{\rho}_*=\frac{a+bz}{1+(z/c)^d}
\end{equation}
and we fit the values of $a,b,c,d$ using a monte carlo method to find a best fit.  This leads to some random fluctuations in the final results due to degeneracies in fits of $a,b,c,d$ to the data however, as we shall see, the overall trend is clear.  We have tried using the median value of the fit rather than the best fit but it doesn't significantly reduce the noise in the final data.  Use of a larger star formation rate data set in future might help in reducing this scatter.  We shall come back to this issue when we discuss the issues with errors that need to be addressed in the future in order for this to be a reliable method of constraining dark energy in section \ref{errors}.

\section{Exact Procedure and Results}

This section describes the procedure for obtaining results.  First we choose a cosmology which has dark energy with a nonstandard equation of state.  The way we parametrise the equation of state is the following:-
\be
w(z)=w_o+\frac{w_az}{1+z}.
\label{eos}
\ee 
We assume that the cosmology is flat and that $\Omega_M=0.276$ which is close to the WMAP value \cite{wmap}.  We also assume a Hubble parameter of $h=0.7$.  Ideally we would like to relax these assuptions and allow these parameters to vary, but the difficulty we will encounter in coming up with a reliable $\chi^2$ statistic for the dimmming of photons in different cosmologies makes this difficult.  We will discuss these issues in section \ref{errors}.

Having chosen a particular cosmology (i.e. a particular pair of $w_o$ and $w_a$), we obtain the comoving distance as a function of redshift as well as the time vs. redshift relationship.  We then apply the appropriate rescaling outlined in equation (\ref{rescale}) to the $\dot{\rho}_*$ data to make up for the different between this particular cosmology and the $\Lambda$CDM cosmology assumed in reference \cite{sfr}.   We then vary the parameters $a,b,c$ and $d$ in order to fit the rescaled star formation rate data using a monte carlo method, basically by using the Metropolis-Hastings algorithm and looking for the best fit value.
\begin{figure}
\begin{center}
\includegraphics[scale=0.35,angle=0]{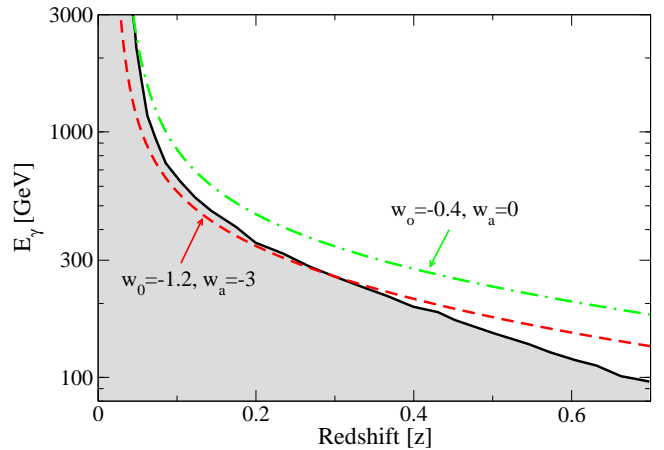}
\caption{\it Figure showing how some cosmologies are seemingly incompatible with observations.  The black solid line is the constraint on the $\tau=1$ gamma ray opacity from \cite{magic}.  The Green dot-dashed line and the red dashed line correspond to two different cosmologies with equations of state indicated in the figure (both are for $h=0.7$ and $\Omega_M=0.276$).  The cosmology corresponding to the Green line is compatible with the gamma ray opacity data while that correspidning to the red line is not.\label{taus}}
\end{center}
\end{figure}
Then we produce the extragalactic background light spectrum throughout the history of the new cosmology using the fitted values of $a,b,c$ and $d$ and the procedure described in section \ref{ebl}.  Next we follow the procedure outlined in section \ref{opac} to work out how far high energy gamma rays can propagate through that newly derived radiation field.  In particular we calculate the position of the line corresponding to the $\tau=1$ opacity in the $E_\gamma - z$ plane (gamma ray energy vs. redshift).  We then compare this line with the line which has been published in the literature from the Cerenkov telescopes.  An example of two such lines, one of which is compatible with the observed dimming and the other which is not can be seen in figure \ref{taus}.

\begin{figure}[t]
\begin{center}
\includegraphics*[viewport=100 120 450 730,scale=0.4,angle=270]{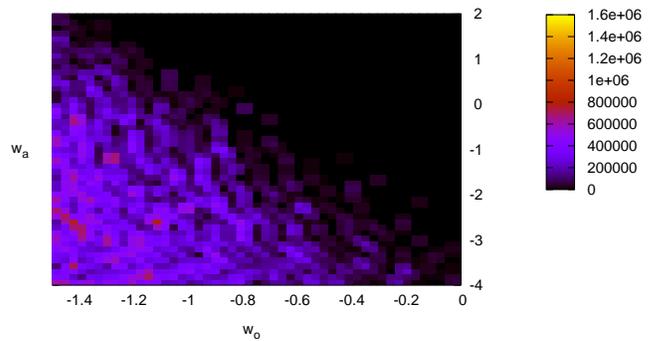}
\caption{\it The value of the overhang parameter $\Psi$ defined in the text for different values of $w_o$ and $w_a$.  It is clear that cosmologies in the lower left part of the diagram are disfavoured.\label{splot}}
\end{center}
\end{figure}

\begin{figure}[t]
\begin{center}
\includegraphics[viewport=70 43 350 200]{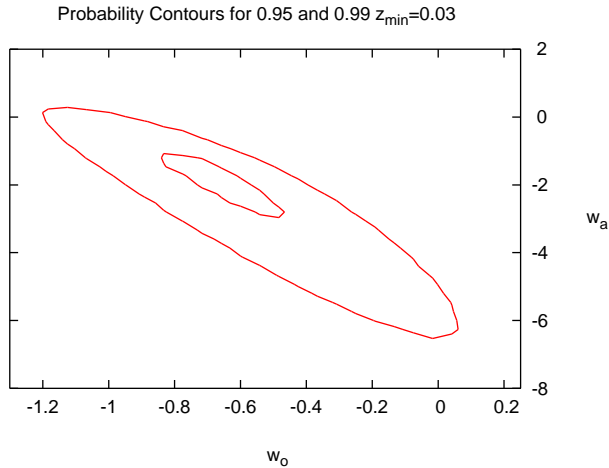}
\caption{\it Constraints on the parameters $w_o$ and $w_a$ as obtained from Supernova Data, Baryon Acoustic Oscillations and Cosmic Microwave Background Data in reference \cite{voidpaper1}.  The two contours correspond to 0.99 and 0.95 times the maximum probability using supernova data from $z=0.03$ upwards.  This should be compared with figure \ref{splot}.\label{normal}}
\end{center}
\end{figure}

These $\tau=1$ lines correspond to a one to one mapping between $E_\gamma$ and $z$ which we will denote $\left.E_\gamma(z)\right|_{\tau=1}$.  We will call the reference limit on the opacity obtained by Magic $f(z)=\left.E_{\gamma MAGIC}(z)\right|_{\tau=1}$ while the function obtained for a particular $w_o-w_a$ cosmology we will call $g(z,w_o,w_a)=\left.E_{\gamma COSMOLOGY}(z)\right|_{\tau=1}$.  We can then define an overhang parameter $\Psi(w_o,w_a)$ as
\be
\Psi(w_o,w_a)=\int H(f(z)-g(z,w_o,w_a))\left(f(z)-g(z,w_o,w_a)\right)^2dz
\ee
where $H(x)$ is the Heavyside step function.  This is simply the square of the area that the $\tau=1$ opacity curve overhangs the reference opacity curve on the $E_\gamma - z$ plane.  We plot this overhang parameter as a function of $w_o$ and $w_a$ in figure \ref{splot}.
This figure shows clearly that cosmologies with low values of $w_0$ and $w_a$ are disfavoured by looking at the gamma ray opacity of such universes.  This is in contrast to the cosmological data, which restricts the values of $w_o$ and $w_a$ to lie inside a diagonal ellipse as can be seen in figure \ref{normal}.  Comparison of figures \ref{splot} and \ref{normal} shows that regions which are not constrained by any cosmological measures are placed under pressure by the gamma ray opacity method outlined in this paper.  The method therefore can therefore in principle regions of parameter space which the rest of the cosmological data set is consistent with.
The current work is a proof of principle for this method.  In the next section we will outline what needs to be done in order for this method to be placed on a stronger footing.

\section{Obtaining Errors and Future Improvements\label{errors}}
In the previous section we saw that there is tension between the observed gamma ray opacity of the Universe and some combinations of the equation of state parameters $w_o$ and $w_a$ which cannot be ruled out with the normal cosmological data.  We also defined an overhang parameter $\Psi(w_o,w_a)$ which measured how baldy each cosmology disagreed with the gamma ray opacity constraints.  In principle it would be great to obtain a mapping between $\Psi$ and a probability, then this probability could be brought together with the other coventional constraints on the dark energy equation of state in a combined analysis.  However this is not yet possible, for a number of reasons that we have until now not mentioned.  In this section we will outline quite a number of these issues which will make up a to-do list which needs to be completed in order for the method outlined in this work to be usefully employed.

In order to turn the $\Psi$ statsictis into a $\chi^2$ statistic we would have to compare the overhang with {\it errors} on the gamma-ray opacity obtained at various redshifts.  The error on the constraint on the gamma-ray opacity provided by Magic is not quoted in their paper because they have not yet been able to calculate it in a robust way.  They obtain the opacity as a function of redshift by looking at the distortions in apparently power law spectra which are interpreted as being due to electron-positron pair production as the photon crosses the intergalactic radiation field.  However, this is an assumption which it is difficult to prove.  Progress into coming up with a more rigorous statistic has been made in a recent paper by the Fermi collaboration working with more photons \cite{fermi}.  Hopefully the advent of the Cerenkov Telescope Array will allow the gamma ray observational community to come up with more robust meausurements of the gamma ray opcity as a function of redshift based upon more statistics and including errors.

The method for calculating the extragalactic background light is full of assumptions, for example, the shape of the initial mass function and its evolution over time.  In our work we have assumed that one has a Salpeter imf throughout the entire history of star formation but this is clearly an over-simplification.  Furthermore we have not taken into account the way that metallicity changes over time and the variation of metallicity in different spatial regions of the Universe (both of which of course have been observed to vary.)  Only by measuring such variations and shapes as a function of redshift, or at least constraining them, then folding these variations and uncertainties into the analysis to understand the uncertainty in the extra-galactic background radiation field would we be able to obtain a reliable error on the dimming in a particular cosmology.

Finally in the future we may be able to collect larger and more reliable data on the star formation histroy with smaller errors.  This would enable us to increase our confidence in our fitting of the star formation rate paremeters $a,b,c,d$, indeed also to test this parametrisation and reduce a lot of the noise in figure \ref{splot}.

\section{Summary and Conclusions}
In this work we have suggested a new way of constraining the equation of state of dark energy.  We have calculated the gamma ray opacity of Universes with different equations of state by rescaling the star formation rate history according to each new cosmology.  Some universes (i.e. pairs of values of $w_o$ and $w_a$ in equation (\ref{eos})) which are not ruled out by the normal cosmological data set would lead to an extragalactic background light in compatible with the observed arrival of photons from distant objects.

We have also outlined the steps which need to be taken to improve this technique and to place it on a firm statistical footing.  This of course involves obtaining new and better data, both from gamma-ray telescopes constraining the opciaty of the Universe better and from  Galactic surveys learning more about the star formation rate of the Universe.

If these hurdles can be overcome it seems that the technique presented in the current work could cut into regions of dark energy equation of state parameter space that cannot yet be probed be other means.

\section*{Acknowledgements}
The author is grateful for funding provided by the UK Science and Technology Facilities Council.

\end{document}